\documentclass[pra,twocolumn,nopacs]{revtex4}

\usepackage{graphicx}
\usepackage{amsmath}
\usepackage{amsfonts}
\usepackage{amssymb}
\usepackage{epsf}
\usepackage{hyperref}

\begin{document}
\title{A two-frequency acousto-optic modulator driver to improve \\the beam pointing stability during intensity ramps}
\author{B. Fr\"ohlich, T.~Lahaye, B.~Kaltenh\"auser, H.~K\"ubler, S. M\"uller, T.~Koch, M.~Fattori and T.~Pfau}
\affiliation{5. Physikalisches Institut Universit\"{a}t Stuttgart,
Pfaffenwaldring 57, 70569 Stuttgart, Germany}
\date{\today}

\begin{abstract}
We report on a scheme to improve the pointing stability of the
first order beam diffracted by an acousto-optic modulator (AOM).
Due to thermal effects inside the crystal, the angular position of
the beam can change by as much as 1~mrad when the radio-frequency
power in the AOM is reduced to decrease the first order beam
intensity. This is done for example to perform forced evaporative
cooling in ultracold atom experiments using far-off-resonant
optical traps. We solve this problem by driving the AOM with two
radio-frequencies $f_1$ and $f_2$. The power of $f_2$ is adjusted
relative to the power of $f_1$ to keep the total power constant.
Using this, the beam displacement is decreased by a factor of
twenty. The method is simple to implement in existing experimental
setups, without any modification of the optics.

\end{abstract}


\maketitle

\section{Introduction}
An important application of acousto-optic modulators (AOMs) is the
control of laser beam intensities. The power of the sound wave
traveling inside the acousto-optic crystal determines the amount
of light that is diffracted out of an incoming laser beam.
However, thermal effects lead to a displacement of the diffracted
beams when the power of the radio-frequency driving the AOM is
changed. The position stability is a critical parameter in many
applications using AOMs, especially for dipole traps formed by
strongly focused, far off-resonant laser beams~\cite{grimm}. Such
traps are playing a major role in atomic physics nowadays, as they
allow for the realization of new experiments, for example the
Bose-Einstein condensation (BEC) of atomic species that cannot be
condensed in magnetic traps such as cesium or
chromium~\cite{csbec}, or the all-optical generation of a
BEC~\cite{allopt}. Particularly in crossed optical dipole traps,
where two beams have to be overlapped on a 10~$\mu$m scale, a
small change of the beam position can have a dramatic effect on
the trap characteristics (frequency and depth), thus causing
severe problems~\cite{newjou}. One way to circumvent them, is to
use a single-mode optical fibre after the AOM, but this cannot be
done for high power lasers, such as $\rm CO_2$ or ytterbium fibre
lasers. In this paper we report on a simple scheme, adaptable to
any AOM, which strongly reduces the beam displacement. The method
is based on driving the AOM with two different radio-frequencies
$f_1$ and $f_2$, and adjusting their relative powers $P_1$ and
$P_2$ so that the total RF power $P=P_1+P_2$ in the AOM is kept
constant~\cite{phd}. This article is organized as follows: After
describing the experimental setup with which we measure the beam
displacement, we present our measurements for AOMs in the 1~$\mu$m
and the 10~$\mu$m wavelength range. In an appendix we show the
details of the electronic circuit we use to adjust $P_2$ relative
to $P_1$ with a single control voltage.

\section{Experimental setup}

\begin{figure}
\includegraphics{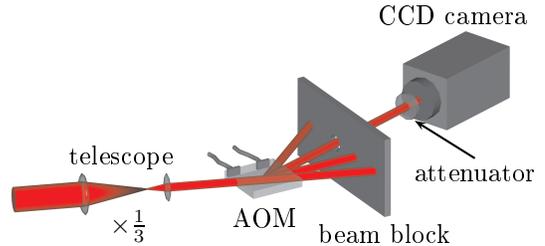}
\caption{(Color online) Setup for measuring the beam displacement
of the AOM using a ${\rm TeO_2}$ crystal. The size of the laser
beam is reduced with a telescope before it enters the AOM. A beam
block after the AOM stops all light except the used beam, which is
attenuated and monitored with a CCD camera. The distance between
the AOM and the camera is 1.4~m.} \label{fig:1}
\end{figure}

We test the two-frequency method with two AOM models that use
different acousto-optic crystals to modulate the light. The setup
for measuring the beam displacement of the first AOM using a
tellurium dioxide (${\rm TeO_2}$) crystal (Crystal Technology
3110-199) is shown in figure~\ref{fig:1}. We use an ytterbium
fiber laser (IPG, model YLR-20-1064-LP-SF) at 1064~nm, with 10~W
output power. The $1/e^2$ beam radius is reduced with a telescope
from initially 2.1~mm to 0.7~mm before going through the AOM.
After the AOM a beam block stops all light except the used beam,
which is attenuated and monitored with a CCD camera. We fit the
images with a 2D-gaussian and record the peak position of the beam
profile. The setup for the second AOM using a germanium (Ge)
crystal (IntraAction Corp. AGM-406B1) is slightly different. We
use a $\rm{CO_2}$ laser (Coherent GEM100L) at 10.6~$\mu$m, with
21~W of power going through the AOM. At a distance of about 3~m we
measure the beam profile in one dimension with a movable pinhole
in front of a power meter. We fit the profile with a gaussian and
record the peak position.

\begin{figure}
\includegraphics{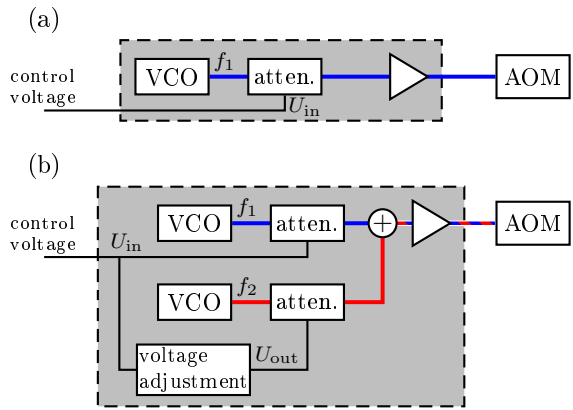}
\caption{(Color online) (a) Normal setup for driving an AOM with
variable RF power. A voltage controlled oscillator (VCO) generates
the radio-frequency $f_1$ (blue line), which is attenuated to a
value given by the control voltage $U_{\rm in}$. The signal is
then amplified before going to the AOM. (b) For the two-frequency
AOM driver we add an extra VCO and attenuator. The additional VCO
generates the second RF signal $f_2$ (red line), whose power is
adjusted relative to $f_1$ to keep the total power in the AOM
constant. This adjustment is done by modifying the control voltage
$U_{\rm in}$ with an electronic circuit (shown in detail in the
appendix). For the TeO$_2$ AOM we use the following Mini-Circuits
components: VCO POS-150, attenuator PAS-3, combiner ZMSC-2-1,
amplifier ZHL-1-2W.} \label{fig:2}
\end{figure}

Figure~\ref{fig:2} shows the modified AOM driver one has to use
for the two-frequency method. To control laser intensities with an
AOM, one has to change the RF power driving it. This can be done
by attenuating a RF signal coming from a voltage controlled
oscillator (VCO) before amplifying it to its final value
(figure~\ref{fig:2}~(a)). The amount of light that is diffracted
out of the incoming beam is then determined by the control voltage
$U_{\rm in}$. For the two-frequency driver we add a second VCO and
attenuator (figure~\ref{fig:2}~(b)) with frequency $f_2$. The two
frequencies $f_1$ and $f_2$ are chosen close enough in order to be
well within the bandwidth of the AOM~\cite{bw}, but far enough to
give a sufficient separation of the two first order beams. We use
$f_1=99$~MHz (resp. 30~MHz) and $f_2=123$~MHz (resp. 50~MHz) for
the TeO$_2$ (resp. Ge) AOM. The power of the frequency $f_2$
generated by the second VCO is adjusted relative to the power of
$f_1$ in order to keep the total power in the AOM constant. To do
this with a single control voltage, $U_{\rm in}$ is modified by an
electronic circuit (see appendix) before it is applied to the
second attenuator. We adjust the transfer function $U_{\rm
out}(U_{\rm in})$ of the circuit to have a constant total RF power
after the signals are added \emph{and amplified}, the latter
condition being crucial to take into account the amplifier
saturation.

\begin{figure}
\includegraphics{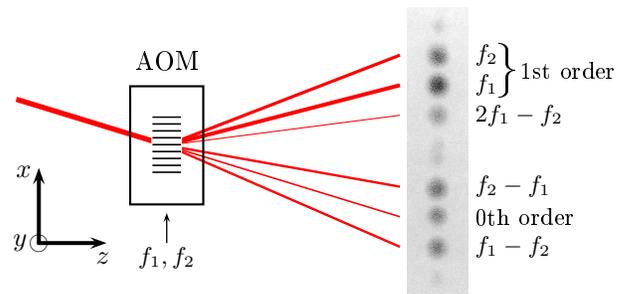}
\caption{(Color online) Schematic of the AOM driven by two
frequencies. The image shows a picture of the laser beam
diffracted by the TeO$_2$ AOM. On the right hand side of the image
the frequency shifts corresponding to the diffracted light are
indicated ($f_1=99$~MHz, $f_2=123$~MHz).} \label{fig:3}
\end{figure}

Laser light going through an AOM driven by two frequencies is
diffracted in many different beams as can be seen in
figure~\ref{fig:3}. The image was taken with the TeO$_2$ AOM at
about equal power of both RF signals. Besides the zeroth order
beam, the first order of both frequencies, as well as second and
even third order beams, which correspond to multiple absorption
and stimulated emission of phonons~\cite{hecht}, can be seen. For
measuring the beam displacement we optimize the angle between the
acoustic wave and the incident laser beam to have the maximum
power in the first order of $f_1$. With full power at this
frequency and none at $f_2$, we achieve diffraction efficiencies
up to 90\%.

\section{Measurements}

\begin{figure}
\includegraphics{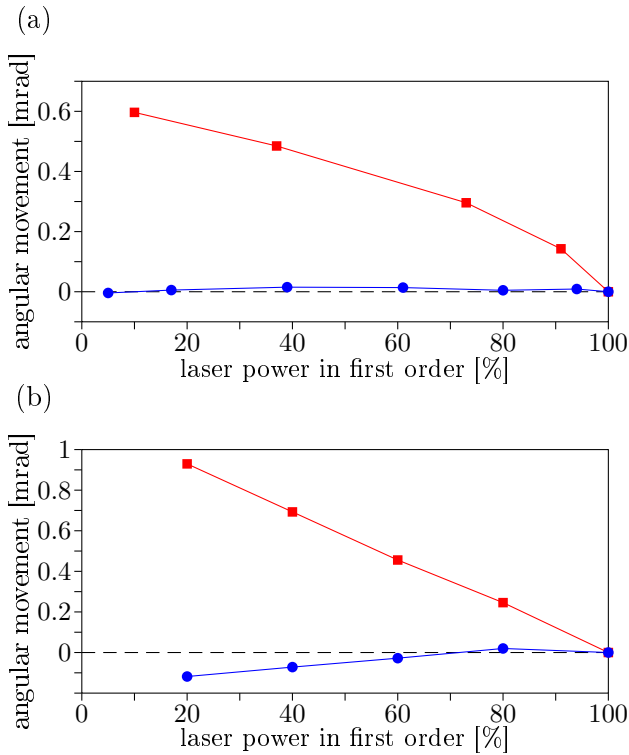}
\caption{(Color online) (a) Measured angular movement of the first
order beam perpendicular to the plane of diffraction ($y$) with
(blue circles) and without the second frequency (red squares) for
the TeO$_2$ AOM. The movement is plotted as a function of the
relative laser power in the first order with respect to its
maximum value. (b) Same measurement for the Ge AOM, measured in
the diffraction plane $x$.} \label{fig:4}
\end{figure}

With the setups described above we measure the position of the
first order beam of $f_1$ at different RF powers for the two AOMs,
with and without the second frequency. In figure~\ref{fig:4} we
plot the angular movement as a function of the laser power in the
first order beam. Figure~\ref{fig:4}~(a) shows the displacement
perpendicular to the plane of diffraction~$y$ for the TeO$_2$ AOM.
The displacement in the plane of diffraction~$x$ (not shown in the
figure) has the same dependence as perpendicular to it, but is
smaller by a factor of three. Adding the second frequency keeps
the beam position almost constant (below 0.03~mrad), whereas
without, a beam displacement of up to 0.6~mrad occurs. A big
improvement is also evident for the Ge AOM
(figure~\ref{fig:4}~(b)), the angular movement is reduced by a
factor of ten. The fact that we are not able to compensate the
displacement as well as with the TeO$_2$ AOM is due to the higher
RF power the AOM is driven with. For maximum diffraction
efficiency the Ge AOM needs 30~W RF power, whereas the TeO$_2$ AOM
needs only 2~W. Another TeO$_2$ AOM that we tested (A-A
Opto-Electronics deflector, model MTS80-A3-1064Ac) uses a sheer
mode acoustic wave and needs only 0.5~W RF power for maximum
diffraction efficiency. Its beam movement is significantly smaller
than for the other AOMs, only up to 0.1~mrad, but still larger
than with the two-frequency method~\cite{deflector}.

To supplement those steady state measurements, we have also
checked for the TeO$_2$ crystal that the suppression of the beam
movement remains good, when the RF power is \emph{dynamically}
ramped down over a timescale of a few seconds, as is done for
forced evaporative cooling of ultracold atoms.

\begin{figure}
\includegraphics{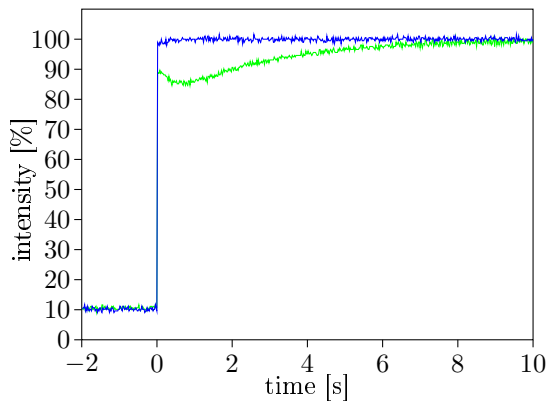}
\caption{(Color online) Time dependence of the laser intensity
when switching the RF power rapidly. Without the second frequency
(green) it takes nearly 10 seconds for the intensity to stabilize
to its steady state value. With the second frequency (blue) there
is only a very small transient effect in the first second.}
\label{fig:5}
\end{figure}

The two-frequency method helps also to stabilize the laser power
$P$ in the first order when switching the RF power rapidly as can
be seen in figure~\ref{fig:5}, which shows the time dependence of
$P(t)$ for the TeO$_2$ AOM. Without the second frequency it takes
about 10 seconds until the steady state value is reached, when
switching the laser power abruptly from 10 to 100\%. The beam
displacement takes place over the same time scale. Only a very
small transient effect in the first second after switching can be
seen, when using the two-frequency method.

In conclusion we have demonstrated a simple method to improve the
pointing stability of a beam diffracted by an AOM when the
intensity is ramped down. The salient advantage of this technique
lies in the fact that only the RF driver has to be modified,
without any modifications of the optics.

\begin{acknowledgements}
We thank C.~S.~Adams for useful discussions and W.~M\"ohrle for
the design of the digital control box. We gratefully acknowledge
the support of the German Science Foundation (DFG) (SFB/TR 21) and
the Landesstiftung Baden-W\"urttemberg. T.~L. acknowledges support
from the European Marie Curie Grant MEIF-CT-2006-038959.
\end{acknowledgements}

\appendix*
\section{Voltage adjustment circuit}

\begin{figure*}
\includegraphics{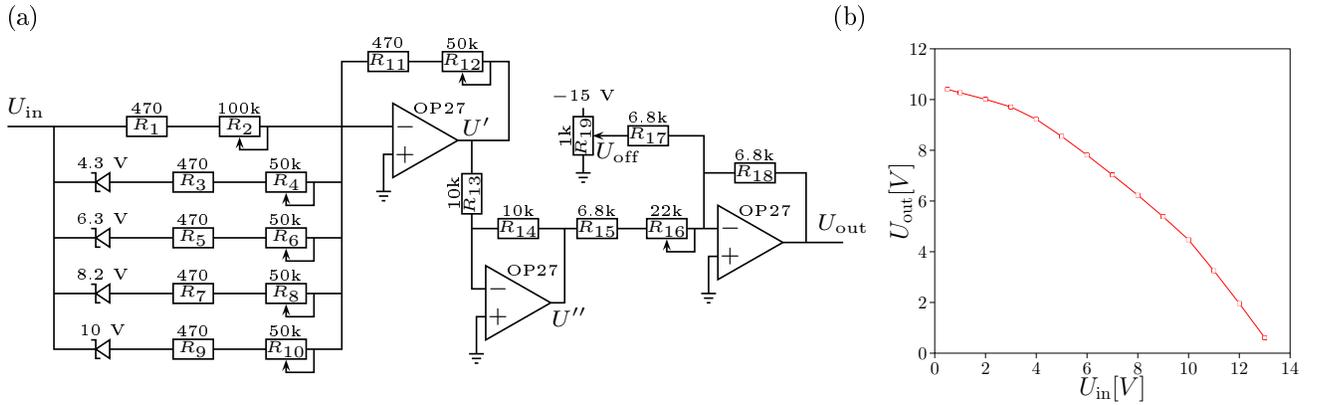}
\caption{(Color online) (a) Schematic of the electronic circuit
for adjusting the control voltage. The gain of the first inverting
amplifier depends on the voltage $U_{\rm in}$ due to the Zener
diodes. The amplified voltage is inverted again before a variable
offset $U_{\rm off}$ is added in the last step. (b) Measured
transfer function of the circuit.} \label{fig:6}
\end{figure*}

In this appendix we present a simple way to realize the voltage
adjustment needed for the two-frequency method
(figure~\ref{fig:2}~(b)). The electronic circuit shown in
figure~\ref{fig:6}~(a) modifies the control voltage $U_{\rm in}$,
so that the total RF power stays constant in the AOM. We measured
the required calibration curve $U_{\rm out}$ as a function of
$U_{\rm in}$, which the circuit approximates by a stepwise linear
function. To do this, we use an inverting amplifier whose gain at
low voltages is given by $-\frac{R_{11}+R_{12}}{R_1+R_2}$.
Parallel to $R_1$ and $R_2$ are other resistors ($R_3, R_4, ...$)
in series with Zener diodes. If $U_{\rm in}$ is larger than the
Zener voltage of one of the diodes it gets conducting and the gain
is increased. For example if $4.3\,{\rm V} \leqslant U_{\rm in}
\leqslant 6.3$~V the gain is increased to
$-\frac{(R_{11}+R_{12})}{(R_1+R_2)
\parallel (R_3+R_4)}$. Thus, each time $U_{\rm in}$ exceeds a
Zener voltage of one of the diodes the gain increases. The
amplified voltage $U'$ is then inverted to $U''$ before in the
last step the voltage $U_{\rm off}$ is added. The potentiometer
$R_{16}$ allows for an extra gain in the last step. We use large
potentiometers for all resistors to have a big flexibility for the
transfer function. In figure~\ref{fig:6}~(b) the measured transfer
function is plotted. With this we are able to keep the total RF
power after amplification constant within 10\%, which is enough to
strongly reduce the beam displacement. For the setup using the Ge
AOM we use a more complex control box, which digitizes $U_{\rm
in}$ with an analog-to-digital converter and then generates the
output voltage $U_{\rm out}$ according to a conversion table
written in an EPROM.


\begin{thebibliography}{99}
\bibitem{bw}
We obtain the AOM bandwidth by measuring the reflected power as a
function of the radio frequency using a directional coupler
(Mini-Circuits ZDC-10-1).
\bibitem{deflector}
One drawback of the acousto-optic deflector is that the sound
velocity for the sheer mode in TeO$_2$ is significantly smaller
(by a factor of five) than for the longitudinal mode, leading to
longer rise times. Using the two-frequency method with a
longitudinal mode AOM allows one to keep fast rise times.
\end{thebibliography}
\end{document}